\documentclass[a4paper,fleqn,usenatbib]{mnras}
\usepackage{graphicx}
\usepackage{graphics}
\usepackage{amsmath, amssymb}
\usepackage{multirow}
\usepackage{color}
\usepackage{xcolor}
\usepackage{bm}		
\usepackage[normalem]{ulem}

\usepackage{float}

\def\kms{{\rm\,km\,s^{-1}}}
\def\kmskpc{{\rm\,km\, \,s^{-1} \, {kpc}^{-1}}}

\def\deg{{^\circ}}

\def\kpc{{\rm kpc}}

\def\mathnew{\mathsurround=0pt}   
\def\simov#1#2{\lower .5pt\vbox{\baselineskip0pt  
    \lineskip-.5pt\ialign{$\mathnew#1\hfil##\hfil$\crcr#2\crcr\sim\crcr}}}

\def\'#1{\ifx#1i{\accent"13\i}\else{\accent"13#1}\fi}

\def\aap{{A\&A}}

\def\araa{{ARA\&A}}
\def\aj{{AJ}}
\def\apj{{ApJ}}
\def\apjl{{ApJL}}
\def\nat{{Nature}}
\def\mnras{{MNRAS}}
\def\apjs{{ApJS}}

\def\rmxaa{{Rev. Mex. Astron. Astrofis.}}
\def\pasp{{PASP}}
\def\msais{{MSAIS}}

\def\nbodysixtt{{\sc{nbody6tt}}}
\def\marms{M_{\rm arms}}
\def\mdisc{M_{\rm disc}}

\newcommand{\rg}{R_{\rm G}}
\newcommand{\vg}{V_{\rm G}}
\newcommand{\feh}{{\rm [Fe/H]}}
\newcommand{\vesc}{v_{\rm esc}}
\newcommand{\rc}{r_{\rm c}}
\newcommand{\rh}{r_{\rm h}}
\newcommand{\rhoism}{\rho_{\rm gas}}
\newcommand{\sech}{{\rm sech}}
\newcommand{\sigmaz}{\sigma_Z}
\newcommand{\taugmc}{\tau_{\rm GMC}}

\newcommand{\rhoh}{\rho_{\rm h}}
\newcommand{\msun}{{\rm M}_\odot}
\newcommand{\dr}{{\rm d}}
\newcommand{\pc}{{\rm pc}}
\newcommand{\gyr}{{\rm Gyr}}
\newcommand{\ggmc}{\gamma_{\rm GMC}}
\newcommand{\Sigmagmc}{\Sigma_{\rm GMC}}
\newcommand{\sigmagmc}{\sigma_{\rm rel}}
  
  \title[Origin and evolution of the open cluster NGC~6791]{New insights in the origin and evolution of the old, metal-rich open cluster NGC~6791}

\author[Martinez-Medina et al. 2017]{Luis A. Martinez-Medina$^1$\thanks{Contact e-mail:
    \href{mailto:lamartinez@astro.unam.mx}{lamartinez@astro.unam.mx}}, Mark Gieles$^2$, Barbara Pichardo$^1$, Antonio Peimbert$^1$
   \\ $^1$ Instituto de Astronom\'ia,
  Universidad Nacional Aut\'onoma de M\'exico, A.P. 70--264, 04510,
  M\'exico, CDMX, M\'exico\\
  $^2$ Department of Physics, University of Surrey, Guildford, GU2 7XH, UK}

\begin{document}
\label{firstpage}
\pagerange{\pageref{firstpage}--\pageref{lastpage}}
\maketitle

\begin{abstract}
NGC~6791 is one of the most studied open clusters, it is massive
($\sim5000\,\msun$), located at the solar circle, old ($\sim8\,$Gyr)
and yet the most metal-rich cluster ($\feh\simeq0.4$) known in the
Milky Way. By performing an orbital analysis within a Galactic model
including spiral arms and a bar, we found that it is plausible that
NGC~6791 formed in the inner thin disc or in the bulge, and later
displaced by radial migration to its current orbit. We apply different
tools to simulate NGC~6791, including direct $N$-body summation in
time-varying potentials, to test its survivability when going through
different Galactic environments. In order to survive the 8 Gyr journey
moving on a migrating orbit, NGC~6791 must have been more massive,
$M_0 \geq 5\times10^4 M_{\odot}$, when formed.  We find independent
confirmation of this initial mass in the stellar mass function, which
is observed to be flat; this can only be explained if the average
tidal field strength experienced by the cluster is stronger than what
it is at its current orbit. Therefore, the birth place and journeys
of NGC~6791 are imprinted in its chemical composition, in its mass
loss, and in its flat stellar mass function, supporting its origin in
the inner thin disc or in the bulge.
\end{abstract}                
 


\section{Introduction} \label{sec:intro}

Open clusters form in the thin molecular gas disc of the
Galaxy \citep{2008ApJ...685L.125D}, within a galactocentric distance
of $\sim10$ kpc and closer than 180 pc away from the disc plane
\citep{2002A&A...389..871D,2012AstL...38..519G}. No more than 10\%
survive their emergence from molecular clouds as bound systems
\citep{2003ARA&A..41...57L}. They range in age between a few million
years and approximately 10 Gyr, and are currently located between 5
and 20 kpc from the Galactic centre
\citep{1970ApJ...160..811F,1993ApJ...406..501N,1995ApJ...447L..95K,2002ApJ...581..258F,2002ASPC..273....7V,2006ApJ...643.1011P,2012MNRAS.419.1860D,2014ApJ...793..110M,2014A&A...564L...9R}. From
the estimated $10^5$  open clusters existing today
\citep{2010ARA&A..48..431P,2006A&A...445..545P}, we only know a few
thousand due to large amounts of reddening and crowding, especially
towards the Galactic centre
\citep{2002ASPC..273....7V,1970ApJ...160..811F}. Only a few clusters
are near enough to be well studied and portrayed.

From all open clusters in the solar neighbourhood, NGC~6791 is
probably the most intriguing, because of its remarkable orbital and
physical characteristics. With a distance of 4 kpc from the Sun, this
cluster is located close to the solar circle at 8 kpc from the
Galactic centre and 0.8 kpc above the plane. It is an 8 Gyr old
system, which makes it the oldest open cluster known
\citep{2012A&A...543A.106B,2010Natur.465..194G,2008A&A...492..171G,2005AJ....130..626K,1965ApJ...142..655K},
and the most metallic with $[{\rm Fe/H}] \sim +0.40$
\citep{2006ApJ...643.1151C,2006ApJ...642..462G,2006ApJ...646..499O}. NGC~6791
is also one of the most massive open clusters with $M \sim 5000
M_\odot$ \citep{2011ApJ...733L...1P}; this means that the cluster
either traveled through the Galaxy on a relatively quiet orbit or it
was born with a much larger initial mass. In particular, due to
  its current mass and height above the Galactic plane, NGC~6791 seems
  to have been formed as a cluster intermediate between an open and a
  globular cluster \citep{1965ApJ...142..655K}.


The initial mass parameter is difficult to determine since it depends
not only on the environment where it was born but also depends
strongly on the characteristics of the orbit it followed since its
formation (e.g., encounters with the spiral arms
\citep{2007MNRAS.376..809G}, interactions with the bulge and disc
tidal fields), as well as on its intrinsic stellar evolution and
two-body relaxation. Regarding the calculation of NGC~6791 initial
mass, work has been done based on observations combined with
theoretical work; the first to estimate the total mass of the cluster
was \citet{1965ApJ...142..655K}, who found a mass of $\sim
3700\,$M$_\odot$, considering only stars with $m_V < 20$. Later on
\citet{1992AcA....42...29K}, confirm the observations but suggest that
a significant fraction of the mass lies in fainter stars. More
recently, \citet{2015MNRAS.449.1811D} derive an initial mass for the
cluster in the range $M= (1.5-4) \times 10^5$ M$_\odot$, using an
expression for the mass evolution of clusters evolving in the Milky
Way disc from \citet{2005A&A...441..117L}.

Several numerical efforts have been done in the past few years to
obtain the orbit of this cluster, which in turn could provide
information on its origin within the Galaxy. Some examples are:
\citet{2006A&A...460L..27B}, who obtain an orbit of approximately 10
kpc of average radius and an eccentricity of 0.5;
\citet{2009MNRAS.399.2146W}, who obtain a similar radius but with an
eccentricity of 0.3 (these two studies were simulated in simple
axisymmetric potentials); or \citet{2006ApJ...643.1151C}, who find a
distance beyond 20 kpc with an eccentricity of 0.59. 

The most recent and complete calculation of NGC~6791 orbit was
  performed by \citet{2012A&A...541A..64J}. In that work the authors
  present a numerical study of the orbit that includes the combined
  effect of the bar and spiral arms in a Galactic disc and halo
  potential. However, the non-axisymmetric part of the model employed
  by the authors for the spiral arms \citep{2002ApJS..142..261C}, for
  instance, does not represent any known mass distribution, i.e., this
  model for the spiral arms is a local approximation (as it is based
  on the addition of sine and cosine functions), this means it is not
  based on a density distribution but it is instead a mathematical ad
  hoc approximation; this type of spiral arm models are intrinsically
  smooth and symmetrical functions that might underestimate the effect
  of spiral arms on the general gas and stellar dynamics.


Alternatively, the peculiar orbit and high $\feh$ of NGC~6791 may
point at an extragalactic origin: in that scenario the cluster, as it
is currently, could be a remnant of an initially more massive cluster
that experienced severe disruption by the MW tides, however
\citet{2017ApJ...842...49L,2015ApJ...798L..41C,2006ApJ...643.1151C},
find a small chemical abundance spread, unlike the one normally seen
in the local group dwarf galaxies \citep{1998ARA&A..36..435M}.

Could an open cluster with such a peculiar combination of parameters
have formed within the Milky Way? (i.e., age, high metallicity, mass,
altitude over the Galactic plane and current galactocentric
distance). Since stars with similar metallicity to the ones within
this cluster have been found in the Galactic bulge
\citep{2013A&A...549A.147B}, \citet{2012A&A...541A..64J} suggested the
possibility that the cluster originated in the bulge, where star
formation was rapid and efficient enough to generate it, however they
concluded this scenario had a very low probability. Of course,
determining the Galactic orbit and survival of the cluster NGC~6791
poses a challenge for several reasons.  Firstly, to restrict the orbit
one depends on the quality of kinematic data. Secondly, a realistic
model of the Milky Way is required, since the orbital behaviour
depends on the details of the spiral arms and bar
representation. Additionally, a good modeling of the internal dynamics
due to stellar evolution, two-body relaxation, and the interaction
with the Galactic tides is key when studying the cluster's disruption
along its path through the Galaxy.

In this work, we explore possible locations for the formation of
NGC~6791, improving previous determinations by including, among other
things, a more detailed model of the MW Galaxy, as well as a more
rigorous criteria in the selection of the orbits. Unlike other
determinations \citep[e.g.][]{2012A&A...541A..64J}, we show there is a
good probability that NGC~6791 was born in the inner thin disc or in
the Galactic bulge region, where an open cluster of such mass and
metallicity is likely to form; we explain in this paper the reasons
for the different results between these authors and our results. We
also present two studies on the survivability of NGC~6791, one based
on a self-consistent field technique and other based on direct
$N$-body models.

This paper is organized as follows. In Section \ref{model}, the
Galactic model, the cluster´s model, the initial conditions, and the
methodology employed are presented. Section \ref{analysis} presents
the orbital analysis and radial migration scenario. Section
\ref{sec:survival} shows the studies of likelihood of survival for
NGC~6791 through different Galactic environments. Section
\ref{sec:Nbody} shows our $N$-body computations of NGC~6791. A brief
discussion is presented in Section \ref{discussion}. Finally, in
Section \ref{conclusions}, we present our conclusions.

\section{Galactic mass model}
\label{model}
For the Galactic model we employ a three dimensional potential with
disc, halo, and rotating spiral-arm and bar components, that fits the
structural and dynamical parameters to the best we know of the recent
knowledge of the Galaxy. We present a summary of the model; a full
description of this potential and its updated parameters appears in a
score of papers
\citep[]{2015ApJ...802..109M,2017MNRAS.468.3615M,PMME03,Pichardo2004,2012AJ....143...73P,2015MNRAS.451..705M}. We
chose to use this Galactic model over the use of the more
sophisticated $N$-body simulations because they are not applicable to
reach the objectives of this work: firstly, because our model is
completely adjustable to better simulate what we know of the MW
(contrary to $N$-body simulations).  Secondly, because it is
significantly faster and therefore more suitable for statistical
studies like the ones presented in this paper. Finally, this type of
models allows us to study in detail orbital behaviour without the
resolution problems of $N$-body simulations.

The model is constituted by an axisymmetric background potential and a
non-axisymmetric one. The axisymmetric potential consists of a
Miyamoto-Nagai disc \citep{1975PASJ...27..533M} with a vertical height
of 250 pc, a Miyamoto-Nagai spherical bulge, and dark matter spherical
halo (based on the potential of \citealt{AS91}). The non-axisymmetric
part includes a three dimensional spiral arms potential and a bar. The
Galactic potential is scaled to the Sun's galactocentric distance, 8.5
kpc, and the local rotation velocity, 220 km s$^{-1}$.

Regarding the non-axisymmetric part of the Galaxy, observations in
infrared bands such as those of the COBE/DIRBE K-band and the infrared
Spitzer/GLIMPSE survey, seem to show that two of the observed arms
are dominant \citep{2001ApJ...556..181D,CHBet09}. Additionally, based
on theoretical work that has shown that two (or more) additional
gaseous arms can  form (without increasing the stellar surface
density) as a response to a two-armed dominant pattern
\citep{2004MNRAS.350L..47M,2015MNRAS.451.2922P}, in this work we will
adopt a two-armed structure for the spiral arms, simulating the main
Galactic spiral arms based on the Spitzer/GLIMPSE database
\citep{BCHet05,CHBet09}. For the spiral arms potential, we employ a
model formed by a bisymmetric three-dimensional density distribution
built of individual inhomogeneous oblate spheroids \citep[PERLAS
  model][]{PMME03} that are placed as bricks in a building within a
logarithmic spiral locus. The density falls exponentially along the
arms. The total mass of the spiral arms taken in these experiments is
$4.28 \times 10^9$ M$_{\sun}$, that corresponds to a mass ratio of
$\marms/\mdisc$ = 0.05. Finally, for the angular velocity, we employ a
value of ${\Omega}_S = 20-28 \kmskpc$, motivated by different
observational and theoretical methods \citep{2017MNRAS.468.3615M,G11}. 

For the bar potential we selected the triaxial inhomogeneous ellipsoid
of \citet{Pichardo2004}; this is a superposition of homogeneous
ellipsoids made to reach a smooth density fall that approximates the
density fall fitted by \citet{1998ApJ...492..495F} from the COBE/DIRBE
observations of the Galactic centre. The total mass of the bar is $1.4
\times 10^{10}$ M$_{\sun}$, within the observational limits
\citep[e.g.,][]{K92,Z94,DAH95,B95,SUet97,WS99,Aet14}. Finally, the
angular speed is observationally set within the range: $\Omega =
45-55$ km s$^{-1}$ kpc$^{-1}$ \citep[and references therein]{2009ApJ...700L..78A,G11}.

For a detailed list of the parameters adopted here to simulate the
Galaxy, see in particular Section 2 of
\citet{2017MNRAS.468.3615M}. Brief descriptions and remainders of the
parameters employed on each simulation of this study are presented
across the paper.

\section{Orbital analysis / Radial Migration scenario}
\label{analysis}

As suggested by its high $\feh$, the location of formation of NGC~6791,
should be far away from its current galactocentric position: it may
have formed in the inner thin disc or in the bulge. However, placing
its birth near the Galactic centre could be a major challenge:
it requires a dynamical mechanism  to
move the cluster from the inner Galaxy to its current position at the
solar circle; in addition, we need to assess the probability for this
mechanism to happen.

The orbit of the cluster can be moved to larger radii by increasing
its angular momentum. It is known that radial migration occurs when a
star exchanges angular momentum with the non-axisymmetric structures
in the disc, changing permanently the guiding radius of the
orbit. Radial migration is expected to be more efficient when it is
driven by a transient spiral pattern \citep{2002MNRAS.336..785S},
however, simulations show that it can also be triggered by the
combined presence of a bar and spiral arms, which can displace orbits
radially without a significant increase in its eccentricity \citep{2016MNRAS.463..459M, 2016MNRAS.461.3835M, 2017MNRAS.468.3615M}.

In this section we explore whether radial migration occurs in our MW
mass model, as well as which combinations, of dynamical parameters for
the bar and spirals, are more efficient in bringing orbits from the
inner disc to the solar circle.

In order to perform this orbital analysis, we first construct four MW
mass models, as described in the previous section; these models differ
in the pattern speeds of the bar and spirals. Table \ref{tab:models}
indicates the combination of pattern speeds used for each MW mass
model. We then populate each model with a particle disc
  distribution composed of 500,000 test particles, and finally
  integrate their orbits for 8.5 Gyr. From these orbits, we pick
those that at the end of the simulation have similar positions, proper
motions, and radial velocities (within $2\sigma$) to the current ones
for NGC~6791.

\begin{table}
\centering
\caption{Galactic mass models constructed for the orbital analysis.}
\label{tab:models}
\begin{tabular}{lcc}
 \hline
 \hline
Model & \multicolumn{2}{c}{\it Pattern speeds (km s$^{-1}$ kpc$^{-1}$)} \\
	& {\it Bar} & {\it Spiral Arms}\\
 \hline
A         & 45            & 20    \\
B         & 45            & 28    \\
C         & 55            & 20    \\
D         & 55            & 28   \\
 \hline
\end{tabular}
\end{table}

The coordinates assumed for NGC~6791 are computed as follows: from the
current heliocentric equatorial coordinates, proper motion, parallax,
and radial velocity of NGC~6791 (see Table \ref{tab:coords}), and the
formula by \citet[][updated to the International Celestial Reference
  System]{1987AJ.....93..864J}, we obtain the kinematics in
coordinates centred on the Sun; the transformation returns, in a
right-handed cylindrical coordinate system, the radial, rotational,
and vertical components of the velocity (U, V, W), that are positive
in the direction of the GC, Galactic rotation, and north Galactic
pole, respectively.  Then, by placing the Sun at ($X,Y,Z$)$_{\odot}$ =
(8.5,0.0,0.025)\kpc, adopting the Sun's velocity with respect to the
LSR ($U,V,W$)$_{\odot}$ = (11.1, 12.24, 7.25) km s$^{-1}$
\citep{2010MNRAS.403.1829S}, and a $V_{\rm LSR}$ = 220 km s$^{-1}$, we
compute the translation from a coordinate frame centred on the Sun to
a cylindrical coordinate frame centred on the Galaxy (Table
\ref{tab:coords}).

\begin{table}
\centering
\caption{Adopted observational data for NGC~6791}
\label{tab:coords}
\begin{tabular}{lcc}
 \hline
 \hline
\multicolumn{3}{c}{\it Equatorial coordinates, radial velocity, PM, distance to the Sun.} \\
\hline
	& {\it Value} & Reference\\
 \hline
$\alpha$                           & 290.22083$\deg$                  & (1)   \\
$\delta$                            & 37.77167$\deg$                    & (1)   \\
$v_r$                                & -47.1 $\pm$ 0.7 km s$^{-1}$     & (2)   \\
$\mu_{\alpha}\cos\delta$  & -0.57 $\pm$ 0.13 mas yr$^{-1}$& (2)   \\
$\mu_{\delta}$                  & -2.45 $\pm$ 0.12 mas yr$^{-1}$& (2)   \\
$d_{\odot}$                      & 4.01 $\pm$ 0.14 kpc              & (3)   \\
 \hline
 \hline
\multicolumn{3}{c}{\it Galactocentric coordinates}\\
 \hline
 	&\multicolumn{2}{c}{{\it Value}}  \\
\hline
$R_{\rm GC}$                        &\multicolumn{2}{c}{8.05 $\pm$ 0.36 kpc}             \\
Z                                   & \multicolumn{2}{c}{0.78 $\pm$ 0.0315 kpc}         \\
U                                  &\multicolumn{2}{c}{ 39.8 $\pm$ 3.9 km s$^{-1}$}     \\
V                                  & \multicolumn{2}{c}{174.74 $\pm$ 3.3 km s$^{-1}$}   \\
W                                  & \multicolumn{2}{c}{-12.13 $\pm$ 3.0 km s$^{-1}$}    \\
 \hline
 \multicolumn{3}{l}{(1) WEBDA; (2) \citet{2006A&A...460L..27B}; (3) \citet{2011A&A...525A...2B}}
\end{tabular}
\end{table}

We note that \citet{2012A&A...541A..64J} use different velocities;
unfortunately, they omitted to rotate $U$ and $V$ by $\sim28\deg$ to
obtain the correct Cartesian velocities (since the position vector of
NGC~6791 is not aligned with the $X$-axis), before computing their
orbits. This missing step in the calculations by J{\'{\i}}lkov{\'a} et al. leads
to a misrepresentation of the orbit of NGC~6791, that in turn leads to
underestimate the efficiency of the migration mechanism (in Private
communication with J{\'{\i}}lkov{\'a} et al., after fixing the omitted velocity
rotation, their re-calculated efficiency of displacing the orbit of NGC~6791 to its
current position increases by an order of magnitude).

In addition, although these authors start from very
precise proper motion and radial velocity, these values are only used
to compute an average peri-galacticon and apo-galacticon for the
present orbit of NGC 6791, which in turn are the ones used for their
orbital analysis; in this manner, the precise information of the
observed kinematics of the cluster is diluted, making it impossible to
identify any possible chaotic behaviour in the orbit
of the cluster.

\subsection{Migrating orbits}
\label{orbits}

Once we have the current galactocentric position and velocities for
NGC~6791, we analyse our integrated orbits and pick the ones with all
their parameters similar to those of Table \ref{tab:coords}.  First we
assume a cluster age of $8.0 \pm 0.5\,$Gyr, so we look for orbits in
the $[7.5, 8.5]\,$Gyr age interval. From those orbits we select the
ones that meet the criteria of having $(R,Z)\in (R\pm 2\sigma_R,Z\pm
2\sigma_Z)_{\rm NGC~6791}$ and $(U,V,W)\in(U\pm 2\sigma_U, V\pm
2\sigma_V, W\pm 2\sigma_W)_{\rm NGC~6791}$, where the observational
uncertainty is taken as the dispersion $\sigma$.

The complete procedure for this orbital selection is as follows: for
the entire set of orbits first we constrain on R and Z; then, for
those orbits the rotation velocity is computed $V = (Y/R)V_X -
(X/R)V_Y$ and we select the ones with $V \approx V_{NGC6791}$;
similarly, we further constrain on radial velocity $U = -(X/R)V_X -
(Y/R)V_Y$ with the condition $U \approx U_{NGC6791}$. Finally, the
last constraint for the remaining orbits is to have $W = V_Z \approx
W_{NGC6791}$.

\begin{figure}
\begin{center}
\includegraphics[width=9cm]{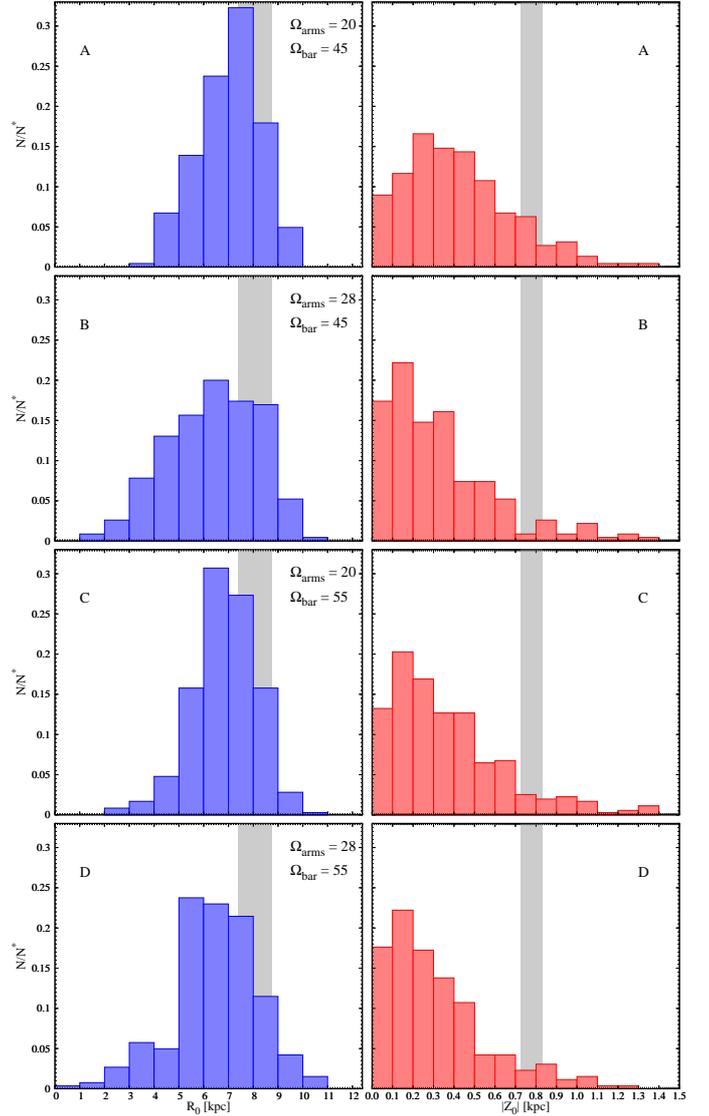}
\end{center}
\caption{Distribution of initial galactocentric radii, $R_0$, and
  initial vertical separation from the plane, $|z_0|$, for orbits that
  meet our criteria (i.e. with kinematics similar to those of
  NGC~6791, see text). The grey bands show the final R and z values of
  those orbits. $N$ is the number of orbits at a given initial radius
  and $N^*$ is the total number of selected orbits. The adopted values
  for $\Omega_{\rm arms}$ and $\Omega_{\rm bar}$ are indicated in
  units of km s$^{-1}$ kpc$^{-1}$.}
\label{f1}
\end{figure}

The result of this analysis, performed in our four Galactic models is
presented in Figure \ref{f1}. The figure shows the initial
galactocentric radius (left column) and initial height above the
Galactic plane (right column) for the orbits that meet the selection
criteria.

The final point of each of these orbits has very similar position and
velocity to the current one of NGC~6791; Figure \ref{f1} shows their
birth positions. The first thing to notice is that, although all
selected orbits end at $8\,$kpc from the Galactic centre, most of them
come from the inner disc; this is a common trend in our four Galactic
models, and all of them show that an orbit like that of NGC~6791
(Table \ref{tab:coords}) is more likely to have come from smaller
radii, with the highest probabilities between $5 < R_0/\kpc < 8 $.

Model B is the one with the widest initial radii distribution, which
implies radial displacements induced over a greater region of the
disc. Also, for this model the mean value of the initial radius is the
lowest, $\left<R_0\right>=6.37 \,$kpc, i.e., this combination of
pattern speeds for bar and spiral arms is more efficient in bringing
orbits from the inner disc to the solar circle.

Additionally, because we are invoking the mechanism of radial
migration to explain the high metallicity of NGC~6791, at odds with
its age and current position, we look for orbits formed within the
specific radial range $3 < R_0/\kpc < 5\,$kpc. Figure \ref{f1} shows
that such orbits exist, and although the probability of finding them
is greater in model B, these orbits exist in all four Galactic models,
which means that they are not restricted to a particular combination
of pattern speeds. 

A final requirement for the orbit of NGC~6791 is to start near the
Galactic plane ($Z\sim0$), because that is where the cold molecular
gas is from which the cluster formed.  Figure \ref{f1} (left column)
shows that most of the displaced orbits start their evolution within
the thin disc, with a significant probability of being formed very
close to the Galactic plane. Again, model B is the most efficient in
bringing orbits from the Galactic plane to altitudes of $\sim800\,$pc
(to match the current vertical position of NGC~6791). Because both the
lifting of the orbits as well as the exchange of angular momentum are
more efficient for orbits interacting more time with the
non-axisymmetric structures of the disc, it is expected for model B,
the one inducing larger radial displacements, to be the one more
efficient in bringing orbits from the thin disc to high altitudes. 

We find 240 orbits out of the
  257,352 that originate between 3 and 5 kpc in our model, that match
  (within 2$\sigma$) the current orbital parameters of NGC~6791. If we
  were interested only in matching the position, the number of orbits
  increases to 3239. This scenario explains naturally its high
  metallicity, in spite of its old age and large Galactocentric radius.

\subsection{Likelihood of observing an anomalous open cluster, like NGC~6791, in the solar neighbourhood}
\label{sec:probability}

To understand the likelihood of the existence of a cluster like
  NGC~6791 in our Galaxy, one should not compare the fraction of
  orbits that match NGC~6791, but rather look for the fraction of
  orbits that would produce an anomalous object similar to NGC~6791,
  and to figure out, for each of these orbits, which fraction of the
  time they would appear to be anomalous.

To do this, we first need to determine the criteria that define
  an orbit as anomalous (in the NGC~6791 sense). The orbits we are
  interested in started between 7.5 and 8.5 Gyr ago, near the galactic
  plane, with a galactocentric radius between 3 and 5 kpc. We consider
  them to produce an ``interesting orbit'' if, at the present time,
  they are far away from the Galactic plane, have migrated outwards at
  least 2 kpc (i.e. have an anomalously high metallicity), and are
  observable from Earth; quantitatively we require these orbits to be
  at least 500 pc away from the galactic plane, with a galactocentric
  radius of at least 7 kpc, but at most 8 kpc away from the Sun.

We explored 257,352 orbits, and for each orbit, we took a hundred
  snapshots (to represent the range of valid ages for the cluster);
  out of these 25,735,200 possible events, we found 61,208 that met
  our criteria; i.e one out of each 420 events would produce such an
  object. This means that, if approximately 420 moderately massive
  clusters were formed between 7.5 and 8.5 Gyr ago, in the Galactic
  plane with a galactocentric radius between 3 and 5 kpc, we would
  expect to observe today an object like NGC~6791 (we will see in
  Section \ref{sec:mass} that we need 420 clusters with an initial
  mass of at least 50,000 solar masses).

As a result of the orbital analysis in this section, we have shown
that it is plausible that NGC~6791 formed in the inner thin disc or
even in the bulge region.

\section{Survivability of NGC~6791 through different Galactic environments}
\label{sec:survival}
In the previous section we have shown that it is possible that
NGC~6791 was born in the inner thin disc or bulge and suffered an
outward radial migration that brought the cluster to its current
position. An important question for this evolution scenario remains:
is the cluster able to survive this? Not only do we need a dynamical
mechanism that displaces the orbit, but also the cluster must manage
to survive its journey from the inner disc to the solar circle.

In Section \ref{analysis}, we found several possible orbits for
NGC~6791; in this section we study whether the cluster can
survive the $\sim8$\,Gyr evolution moving on these orbits, and under
which conditions it can be done.

\subsection{Modeling stellar clusters with a self-consistent field technique}
\label{modeling}

The disruption time depends on the strength of the Galactic tides
(directly related to the type of orbit), initial mass, and size of the
cluster.  Modeling NGC~6791, and analysing its disruption time, for
several combinations of these three factors requires hundreds of
simulations.  Due to the computation time, this is not viable to do
with a direct $N$-body code. Instead, for this part of the work, we
use the technique described in \citet{2017ApJ...834...58M} to model
NGC~6791.  The method assumes that the cluster is spherically
symmetric such that the gravitational potential can be approximated
with that of a \citet{1911MNRAS..71..460P} model, which corresponds to
the zeroth order member in a basis expansion of the gravitational
potential \citep{1972Ap&SS..16..101C, 1973Ap&SS..23...55C,
  1992ApJ...386..375H}. In order to capture mass loss, the criteria to
decide the membership of a given star to the cluster is that the
star's velocity, $v$, must be less than the local escape velocity,
$\vesc$. Those stars with $v>\vesc$ are removed from the cluster,
which will cause the cluster to lose mass. This definition of $\vesc$
does not include the contribution of the tides to the (Jacobi) energy
of the stars, mainly because this is ill-defined when the tides are
time-dependent. \citet{2011MNRAS.418..759R} used $N$-body simulations
of clusters on circular orbits to show that this simple escape
criterion results in a bound mass evolution that is similar to models
in which the tides are included in the definition of $v_{\rm esc}$. In
addition, the radial scale of the Plummer potential evolves with time
by computing, at each time step, the half-mass radius, $\rh$, of the
distribution, which for the Plummer profile is related to the radial
scale-length, $a$, as $\rh \approx 1.3a$. In this way, the potential
of the cluster, although assumed to be represented by a Plummer model,
is evolving with time, with the ability of mimicking expansions and
contractions, as well as the mass loss of the system.

In this method, stars in the cluster do not interact with one another
directly; instead, the contribution of all of them create the global
potential in which the stars orbit. This is an efficient technique
that allows us to approach the self-gravity of the cluster with
minimal computational overhead. By avoiding two-body relaxation, this
technique isolates the effect of tidal interaction of the cluster with
the Galaxy.

\subsection{Disruption time for each orbit as a function of Galactic environment}
\label{sec:disruption}
 
To assess the role that the Galactic environment (i.e., the type of
orbit) played in the evolution of NGC~6791, we model the cluster as
described above and place it on each one of the orbits that meet the
criteria described in Section \ref{orbits}.  The orbits found in our
orbital analysis start in the inner disc, $3<R/{\rm kpc}<5$, a region of
strong tidal fields, which will cause significant mass loss of the
cluster. The mass loss due to tidal interactions  depends on how
long the orbit stays in the inner disc.

\begin{figure}
\begin{center}
\includegraphics[width=8cm]{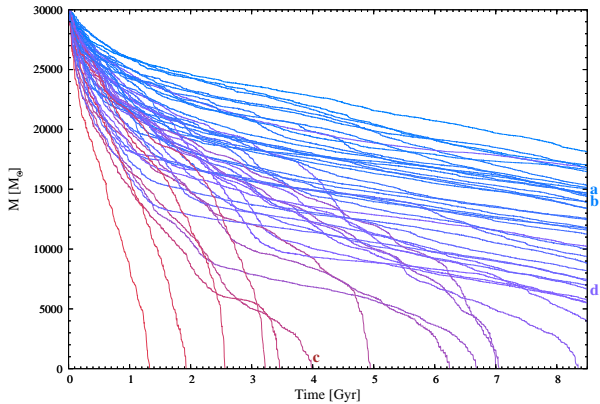}
\end{center}
\caption{Mass evolution for the stellar cluster evolved along the
  orbits integrated with the Galactic model B, and that meet the
  criteria described in section \ref{orbits}. The colour coding
  indicates orbits that pass through strong tidal fields (red), as
  well as those that are less disruptive (blue). The variation of
  $\rg$ of the clusters marked a, b, c and d are shown in
  Fig.~\ref{f3}.}
\label{f2}
\end{figure}

Figure \ref{f2} shows the evolution of the cluster's mass with time,
for the migrating orbits from model B. All clusters are initialized
with the same properties (30,000 stars, each of $1\,\msun$; as well as
an initial $\rh$ of 6 pc), the only difference between the models is the
orbit.

From Figure \ref{f2} we see that the evolutionary paths are diverse,
all orbits end up at the solar circle, $R_{\rm GC} \approx 8\,$kpc,
but before reaching that distance some of them spend more time in the
inner disc. Moreover, some orbits approach the Galactic centre (closer
than their starting positions) before they are displaced to larger
radii. This type of orbits take the cluster to Galactic environments
with strong tidal fields, causing the cluster to rapidly loss mass and
to be disrupted in a short time.  On the other hand, there also exist
orbits that, although starting in the inner disc, are quickly
displaced to larger radii, hence a cluster moving on one of those
orbits will not interact with strong tides for most of its lifetime.

\begin{figure}
\begin{center}
\includegraphics[width=8cm]{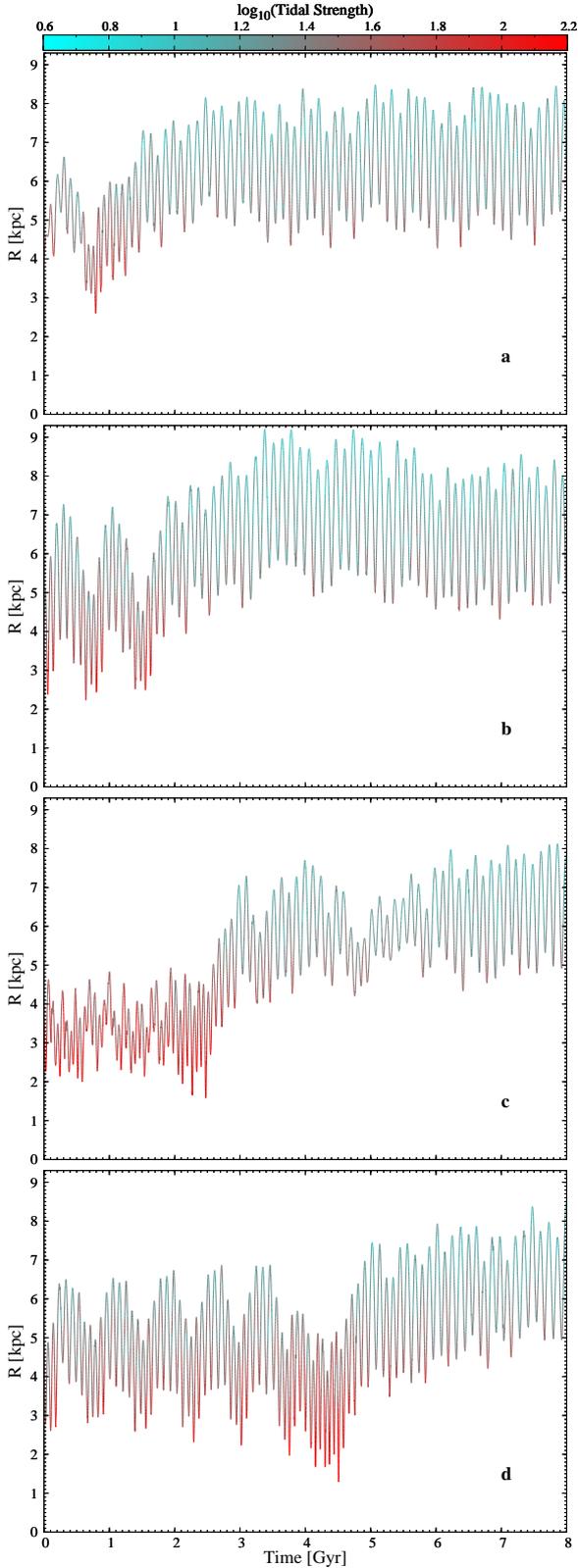}
\end{center}
\caption{Evolution of the cluster's galactocentric radius $R_{\rm GC}$
  for four different orbits that start between 3 and 5 kpc. The colour
  coding indicates the strength of the tides at each point on the
  orbit. Because the initial conditions of the clusters are the same,
  this plot illustrates the importance of the orbital evolution.}
\label{f3}
\end{figure}

To better illustrate how the mass loss of the cluster depends on the
orbit it follows, we show in Figure \ref{f3}  the galactocentric radius
$R_{\rm GC}$ as a function of time for four different orbits (these orbits
are highlighted in Figure \ref{f2}); the colour code indicates the
strength of the tides at every point in the orbit. We choose these
four orbits because they exhibit the different behaviours described
above. All of them start their evolution in the inner region of the
disc, where the tidal field is strong, but the time they spend there
varies from one to another. For the orbits in Figure \ref{f3} the
inhospitality of the Galactic environment increases from top to
bottom. The orbit in the top makes a quick excursion to smaller radii,
but then it is displaced to larger radii in less than 1 Gyr, and stays
there, where the tidal interaction with the Galaxy is less
destructive. The next orbits stay longer periods of time at small
radii, increasing the interaction of the cluster with strong tides,
and hence increasing the mass loss. Notice that the most violent
orbits stay close to the Galactic centre for 3 or even 5 Gyr before
they are displaced to regions of less destructive tides. Hence, the
different behaviour of each orbit determines the mass loss and
evolution of the cluster, as shown in Figure \ref{f2}.

As a consequence of the mass loss dependence on the type of orbit, and
to assure that our models end up with the correct mass after an 8 Gyr
evolution, we need to initialize the simulated clusters with different
initial masses. For the next set of simulations all clusters are
initialized with a $\rh = 6\,$pc, and we impose the condition that at
$t = 8\,$Gyr the remaining mass should be within the range
$4\times10^3 < M/\msun < 6\times10^3$; the first simulation starts
with an initial mass, $M_0 = 3\times10^4\,\msun$; if after 8 Gyr
the remaining mass is not within those values, then the simulation is
repeated with increasing or decreasing $M_0$. By performing this
iteration, we can assign the optimal initial mass for the cluster
according to the type of orbit and Galactic environment.

\begin{figure}
\begin{center}
\includegraphics[width=9cm]{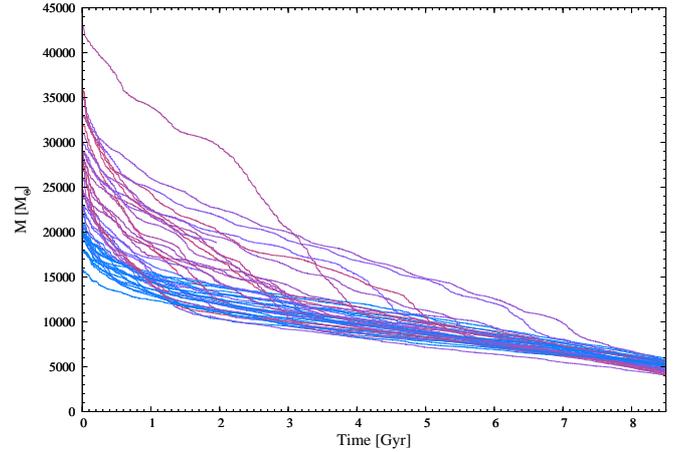}
\end{center}
\caption{Mass evolution for the stellar cluster evolved along the
  orbits integrated with the Galactic model B, and that meet the
  criteria described in section \ref{orbits}. The colour coding
  indicates orbits that pass through strong tidal fields (red), as
  well as those that are less disruptive (blue).}
\label{f4}
\end{figure}

Figure \ref{f4} shows the initial mass exploration, and we can see
that, unlike what happens in Figure \ref{f2}, all clusters survive and
are still massive at $t = 8\,$Gyr. Notice that it is possible to get
all our models to converge to similar evolutionary stages ($M \approx
5\times10^3\,\msun$) as long as we adopt a wide range of initial
masses. As expected, clusters moving on more destructive orbits, i.e.,
interacting with strong tides, will need to be massive at their
formation ($M > 5\times10^4\,\msun$). On the other hand, putting the
cluster on a less violent orbit allows to adopt an initial mass
significantly smaller compared to the one needed for the more violent
orbits ($M > 1.8\times10^4\,\msun$).

Figures \ref{f2}-\ref{f4} illustrate the importance of the Galactic
environment in determining the possible initial mass of NGC~6791. This
implies that, unless we know the exact orbit, this cannot be uniquely
determined.

\subsection{The effect on survivability of initial mass and of initial half-mass radius}

From Figures \ref{f2} and \ref{f4} we notice that to assure the
survival of the cluster, it must have been born significantly more
massive than it is at the present time. Actually, its survivability
also depends on the initial concentration in the distribution of stars
(i.e., the half mass radius of the system at birth). A high
concentration of stars will require a relatively small initial mass,
while a sparse system will require a greater initial mass in order to
assure that the cluster will not be totally disrupted after an 8 Gyr
evolution.

By modeling NGC~6791, as described in Section \ref{modeling}, we
explore here a possible correlation between $M_0$ and the initial
$\rh$ of the cluster. We study all the possible combinations of
$(M_0,\rh)$ within the intervals $6\times10^3 < M_0/\msun < 1.5 \times
10^5$ and $3 < \rh/\pc < 13.5$, these intervals are explored with
increments $\Delta M_0 = 10^3\,\msun$ and $\Delta \rh = 0.5\,$pc. From
this set of simulations, we choose the ones where the mass of the
cluster, after an 8 Gyr evolution, is within $2\times10^3< M/\msun <
8\times10^3$. Here we use the same orbit for all modeled clusters in
order to isolate the role of $M_0$ and $\rh$ in the evolution of the
cluster.

\begin{figure}
\begin{center}
\includegraphics[width=9cm]{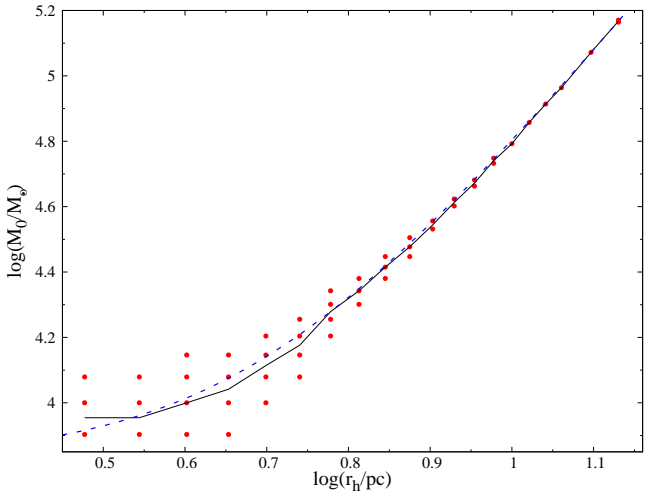}
\end{center}
\caption{Relation between initial mass $M_0$ and initial half mass
  radius $\rh$ (black solid line). Every point corresponds to models
  that after an 8 Gyr evolution still have a remaining mass similar to
  the one of NGC~6791. The data can be adjusted by a cubic relation
  between $M_0$ and $r_h$, indicated here with dashed blue line.}
\label{f5}
\end{figure}

The resulting relation from this exploration of values for $M_0$ and
$\rh$ is shown in Figure \ref{f5}. The dispersion of values is small,
showing a clear correlation between initial mass and initial half mass
radius, which means that these two important properties of the cluster
are not independent. For $log(\rh/\pc)\gtrsim0.8$ the models follow a
line of constant initial half-mass density (i.e. $M\propto
\rh^3$). This can be understood as follows: these models dissolve
under the influence of tidal perturbations (i.e. relaxation and
stellar evolution are not included), which occurs on a time-scale that
is proportional to the cluster density
(e.g. \citealt{1958ApJ...127...17S} and
section~\ref{sec:gmcs}). Because we forced all clusters to survive,
all surviving initial conditions have the same initial density.

These models do not include the effect of stellar evolution and
two-body relaxation, which both tend to expand the cluster and will
therefore speed up the mass-loss. The initial mass of the models in
Fig.~\ref{f4} are therefore lower limits. In the next section we
consider all effects combined with an $N$-body model.

\section{$N$-body models of NGC~6791}
\label{sec:Nbody}

In addition to the properties previously mentioned,
\citet{2015MNRAS.449.1811D} find evidence of tidal distortions in the
outer parts, and provide a King model fit for the inner part of the
cluster. Meanwhile, \citet{2005AJ....130..626K} present an estimate of
the cluster's mass function (MF) (defined as the number of stars in
linear mass bins $N(m)dm$), which is rather flat implying that the
cluster has lost a large number of its low-mass stars (assuming a
universal IMF).

A comparison with these observables gives us the opportunity to better
constrain our models and find the more likely properties of NGC~6791
at its moment of birth. To perform this, a more detailed modeling of
the cluster needs to be done. For this part of the work we employ
\nbodysixtt\, a state-of-the-art direct $N$-body code, a modified
version of the direct $N$-body integrator NBODY6
\citep{2003gnbs.book.....A} optimised for use with Graphics Processing
Units (GPUs) \citep{2012MNRAS.424..545N}, which was specifically
designed for modelling collisional star clusters. It solves pairwise
gravitational interactions between stars in the cluster and includes
synthetic stellar evolution \citep{2000MNRAS.315..543H,
  2002MNRAS.329..897H}. We here use \nbodysixtt\ `Mode A' to include
the tidal field, which relies on user defined tidal tensors which are
read in and from which the tidal forces are applied in the tidal
approximation \citep{2011MNRAS.418..759R}.

As shown in previous sections, the initial mass of the cluster was
possibly considerably larger than it is now, however the method
employed in Section \ref{modeling} does not consider internal
evolution, which is an important factor in the cluster's mass loss
process. Because of that, for the $N$-body modeling of NGC~6791, we
initialize the cluster with a mass greater than the one used in
Section \ref{sec:disruption}. We find the initial mass iteratively. We
adopt a Kroupa initial mass function (IMF) (i.e. double power-law with
logarithmic slope of -1.3 below $0.5\,\msun$ and -2.3 above it), with
star masses between $0.1\,\msun$ and $100\,\msun$, and a $\rh$ of
2.3$\,$pc. Table \ref{tab:nbody} shows the initial set-up of the
cluster, meanwhile the tidal tensor is pre-computed along one of the
orbits from Section \ref{analysis}, it is passed to \nbodysixtt\ as
described in \citet{2011MNRAS.418..759R}. During the simulation,
escapers that reach a distance of two times the tidal radius, $r_{\rm
  tide}$, are removed, with $r_{\rm tide} = 10\rh$.

\begin{table}
\centering
\caption{Setup for the $N$-body model of NGC~6791}
\label{tab:nbody}
\begin{tabular}{lc}
 \hline
 \hline
\multicolumn{2}{c}{Initial values of the $N$-body models.} \\
 \hline
Initial mass, $M_0$                    & $5\times10^4\,\msun$ \\
Number of stars                        & $8\times10^4$  \\
Initial Mass Function                 & Kroupa               \\
Maximum star mass              & $100\,\msun$ \\
Minimum star mass               & $0.1\,\msun$    \\
Average star mass                     & $0.637\,\msun$\\
Half-mass radius                       & $2.3\,$pc   \\
 \hline
\end{tabular}
\end{table}

\subsection{Cluster evolution: mass and density}
\label{sec:mass}

\begin{figure}
\begin{center}
\includegraphics[width=9cm]{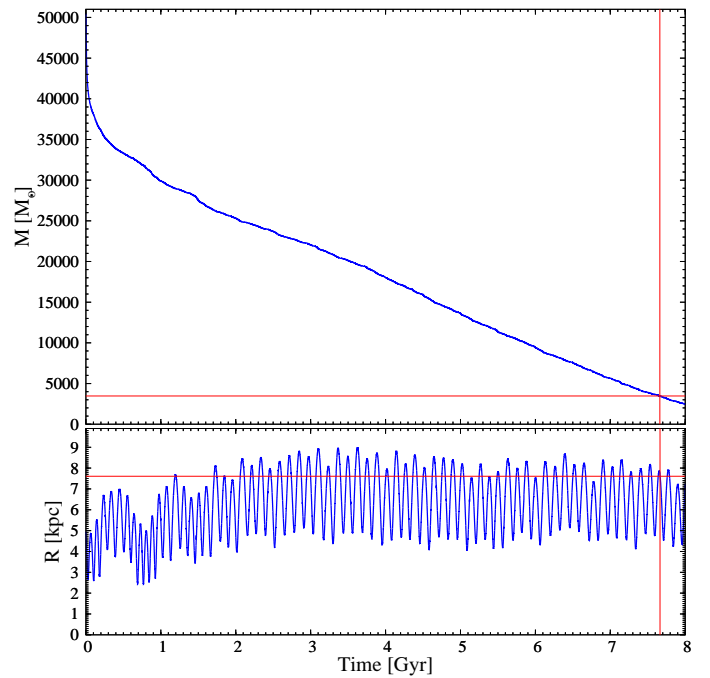}
\end{center}
\caption{Top: cluster stellar mass as a function of time. Bottom:
  evolution of the Galactocentric radius $R_{\rm GC}$ for the orbit of
  the cluster. Red lines indicate the position and mass of the cluster
  at the moment its galactocentric distance, PM, and radial velocity,
  are similar to the current ones of NGC~6791.}
\label{f6}
\end{figure}

The result of the $N$-body modeling of NGC~6791, moving on top of one
of the less violent migrating orbits, is shown in Figure \ref{f6}.The
cluster loses mass along the entire orbit, with significant stripping
at the beginning of its evolution and at the closest approaches to the
Galactic centre, as indicated in the bottom panel of Figure
\ref{f6}. A comparison between both panels shows that the slope of the
mass curve is clearly modified every time the cluster enters a region
of strong tides, this means that the orbit of the cluster is imprinted
in its evolution, and determines its final stage.

{As described in section \ref{orbits}, this particular orbit meets our
  selection criteria at $7.66\,$Gyr, at that moment in the simulation
  the cluster is quite massive, $M = 3473\,\msun$, as indicated in the
  top panel of Figure \ref{f6}. By setting the cluster to follow one
  of the less violent migrating orbits that ends up to be similar to
  the current one for NGC~6791, but also fulfilling the condition of
  ending with a massive cluster after the entire evolution, the
  selected initial mass for this model, $M_0 = 5\times10^4\,\msun$,
  give us a lower limit for the predicted mass of NGC~6791 at its
  moment of birth.}

\begin{figure}
\begin{center}
\includegraphics[width=9cm]{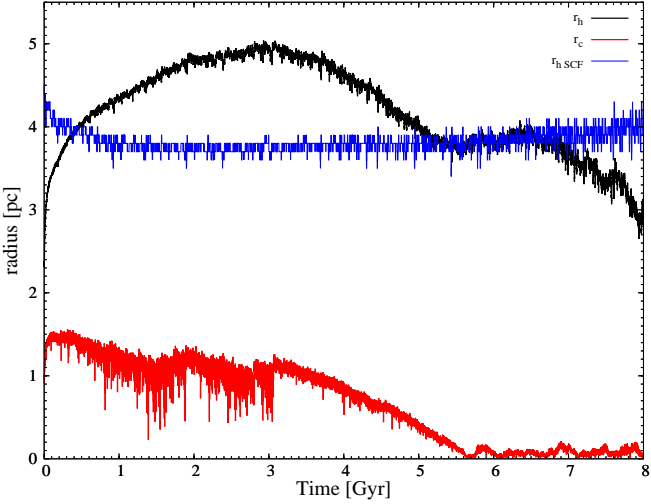}
\end{center}
\caption{Evolution of the half mass radius $\rh$ (black) and core
  radius $\rc$ (red) for the simulation shown in Figure \ref{f6}. For
  comparison, we also indicate the $\rh$ for a model described with
  the SCF method (blue). }
\label{f7}
\end{figure}

Figure \ref{f7} shows the evolution of $\rh$ and core radius,
$\rc$. The cluster undergoes an expansion period of $\sim3\,$Gyr
before reaching approximately $50\%$ of its initial mass, then it
undergoes core-collapse. After $\sim5\,$Gyr, and until $7.6\,$Gyr,
$\rh$ changes rather slowly until its final value. Despite the
time-dependent tidal field, the general behaviour of $\rh(t)$
(i.e. expansion followed by contraction) is similar to the evolution
of clusters in static tides \citep{2011MNRAS.413.2509G}. For
comparison, Figure \ref{f7} also shows the evolution of $\rh$ for a
cluster modeled with the method described in Section
\ref{modeling}. Note that because two-body relaxation is not
considered, the cluster does not undergo the expansion nor the
collapse seen in the $N$-body model.  Because of the lack of
significant expansion, for a cluster modelled with the SCF technique,
the mass loss can be underestimated; this becomes noticeable in high
concentrated models, were the evolution would differ from that of the
$N$-body model shown in Figure \ref{f6}.

\begin{figure}
\begin{center}
\includegraphics[width=9cm]{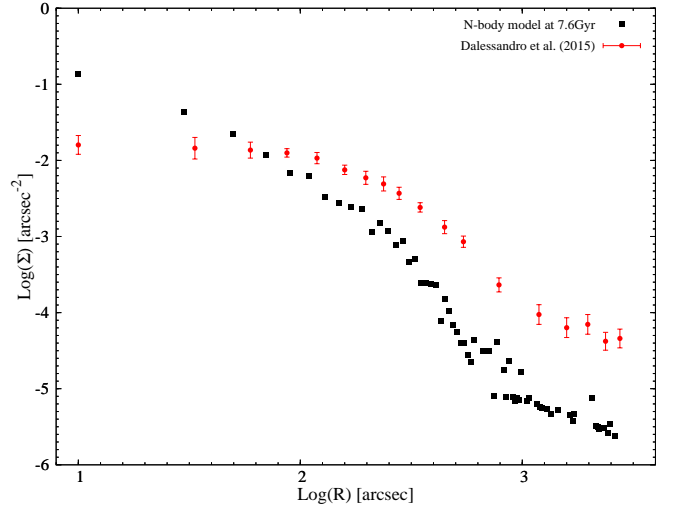}
\end{center}
\caption{Number density profile in projection for the $N$-body model
  at $7.6\,$Gyr as a function of projected radius (black squares),
  compared to the observed number density profile from
  \citet{2015MNRAS.449.1811D}.}
\label{f8}
\end{figure}

Another known property for NGC~6791 is its projected
density. \citet{2015MNRAS.449.1811D} determine the projected density
profile using direct counts of stars along the main sequence in the
magnitude range $18<g'<21.5$. In order to compare with the data, from
the simulation we select stars within the same magnitude range, which
is translated to the mass range $0.73<m/\msun<1.08$. For stars with
those masses, Figure \ref{f8} shows the projected number density
profile of the $N$-body model at $7.6\,$Gyr, as well as the data from
\citet{2015MNRAS.449.1811D}.  The (post-collapse) $N$-body model is
more centrally concentrated than NGC~6791. The number density profile
suggests that NGC~6791 has not yet reached core collapse. The reason
that our $N$-body model has already experienced core collapse could be
due to several reasons: (1) we did not include primordial binary
stars, which delay core collapse \citep[e.g.][]{2011MNRAS.410.2698G};
(2) the stellar-mass black holes received the same super nova kick as
the neutron stars such that few black holes are retained; a black hole
population could also inflate the (visible) core \citep[e.g.][]{
  2004ApJ...608L..25M, 2008MNRAS.386...65M,2016MNRAS.462.2333P}. The
$N$-body model reproduces the `kink' in the number density profile
near $R\sim800\arcsec$, which is due to so-called potential escapers,
which are energetically unbound stars that are associated with the
cluster because of their long escape time \citep{2000MNRAS.318..753F}
and are found predominantly in the outskirts of star clusters
\citep{2010MNRAS.407.2241K, 2017MNRAS.466.3937C}. Just as tidal
streams, the potential escapers are more visible in clusters that are
near dissolution \citep[in terms of remaining mass fraction,
][]{2017arXiv170202543B}, as is the case for NGC\,6791.

\subsection{Stellar mass function}
\label{ssec:mf}

The observed mass function (MF) of a star cluster gives insights into
its IMF and its dynamical evolution. There is an observed correlation
between the MF and the galactocentric distance, being steeper for
clusters more distant from the galactic centre
\citep{1993AJ....105.2148D}; a correlation also found in $N$-body
simulations \citep{1997MNRAS.289..898V,2003MNRAS.340..227B}. Several
studies have shown that for a given IMF, the slope of the present-day
MF is a good proxy of the fraction of the initial cluster mass that
still remains in the cluster \citep[e.g.][]{2003MNRAS.340..227B,
  2010ApJ...708.1598T}. This means that we can use the shape of the MF
and the present-day mass to estimate the initial mass of NGC~6791,
which we can compare to what we find in the $N$-body model.

Using the Hubble Space Telescope (HST), \citet{2005AJ....130..626K}
derive the present-day MF for NGC~6791. They found that the MF is
fairly flat between $0.1\,\msun$ and $1\,\msun$. We follow the MF with
time in the simulation and compare the result to the observations of
King et al.

In order to make a proper comparison with the observed MF, notice that
the HST data of \citet{2005AJ....130..626K} is not covering the entire
cluster. Those observations were made with the Wide Field Channel
(WFC) of the Advance Camera for Surveys (ACS), within a field centred
1.5$^{\prime}$ away from the cluster centre, and the field of view of
the WFC is $3.3^{\prime} \times 3.3^{\prime}$. This means that the
camera covered distances up to 3.15$^{\prime}$ from the cluster
centre. On the other hand, from the King model fit of
\citet{2015MNRAS.449.1811D}, we estimate a 3D $\rh \approx
5^{\prime}$, which means that the MF derived from the HST data is
biased towards the centre of the cluster.  To compare with the MF
reported by \citet{2005AJ....130..626K} we select stars in the
simulation within distances less than 3.15$^{\prime}$, which for an
adopted distance to the cluster of 4.01$\,$kpc (Table
\ref{tab:coords}) translates into a MF for stars within a radius of $r
= 3.66\,$pc.

\begin{figure}
\begin{center}
\includegraphics[width=9cm]{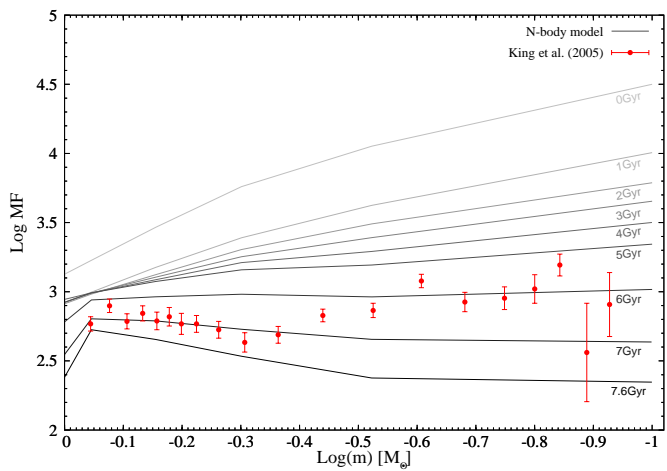}
\end{center}
\caption{Stellar MF for different ages of the cluster.}
\label{f9}
\end{figure}

Figure \ref{f9} shows the MF, for different ages of the cluster. The
MF flattens as the cluster evolves; this happens as a consequence of
the mass loss via the preferential escape of low-mass stars, which is
greater in the first few Gyr, as shown in Figure \ref{f1}. Notice that
the MF from the data is fairly flat (the slope of the MF is significantly smaller than the one of the assumed IMF), meanwhile the MF from the model
reaches a similar slope at around $\sim7\,$Gyr, consistent with the
age of NGC~6791.

A star cluster loses mass through the tidal boundary due to two-body
relaxation and tidal interaction with the Galaxy; both processes
flatten the global MF \citep{1997MNRAS.289..898V,2003MNRAS.340..227B}.
The strength of the tides experienced by the cluster, and hence the
mass loss due to this mechanism, will be larger in the inner regions
of the Galaxy.

With a large set of $N$-body models of multimass clusters evolving in
a Galactic tidal field, \citet{2003MNRAS.340..227B} found a dependence
of the slope of the MF on the mass fraction lost from a cluster. They
parameterize the mean evolution of the slope by

\begin{equation}
\label{slope}
\alpha = \alpha_0 - \exp\left[ 0.67 - 6.19\frac{M}{M_0} - 3.24\left(\frac{M}{M_0}\right)^2\right],
\end{equation}
{where $M$ is the remaining mass of the cluster, $M_0$ its initial
  mass, $\alpha$ is the power-law slope of the MF at any point in the
  evolution of the cluster, while $\alpha_0$ is the slope of the
  IMF. This parameterization allows to infer the initial mass of
  NGC~6791 from its present-day MF and mass.}

Assuming a Kroupa IMF, $\alpha_0 = 1.3$ in the mass range
$0.1<m/\msun<0.5$, while a linear fit to the data from
\citet{2005AJ....130..626K} returns the present-day slope $\alpha =
0.2$. By taking this values as input for equation (\ref{slope}) we
infer a current mass to initial mass ratio of $M/M_0 \approx 0.089$;
and with the current mass of NGC~6791, $M\sim5000\,\msun$
\citep{2011ApJ...733L...1P}, we infer an initial mass of $M_0 =
56180\,\msun$. This estimation is in good agreement with our predicted
value of $M_0 = 5\times10^4\,\msun$ from the $N$-body model placed on
top of a migrating orbit.

Figure \ref{f9} also shows that a flatter IMF and less mass loss would give a similar present-day MF. However, for an assumed IMF, the present-day MF tells us directly how much mass the cluster has lost; but the amount of mass lost can also be estimated by assuming a dynamical history. Because we found agreement between these two mass loss estimations, we conclude that the combination of the assumed IMF and the dynamical history is correct. In this sense we corroborate that, because the slope of the data by  \citet{2005AJ....130..626K} is significantly smaller than the slope of the plausible adopted IMF, the observed MF of NGC~6791 is fairly flat.

Finally, for comparison, we can estimate the slope of the present-day
  MF of a cluster that was born and has always been orbiting at
  8$\,$kpc from the galactic centre. Again here we use the expressions
  by \citet{2003MNRAS.340..227B}: first for the lifetime of a cluster
  moving in a external potential

\begin{equation}
\label{eq:lifetime}
\frac{T_{\rm diss}}{\rm Myr} = 1.91\left[ \frac{N}{\ln(0.02N)}\right]
^{0.75}\frac{\rg}{\kpc}\left(\frac{V_{\rm G}}{220\,{\rm km}\,{\rm
    s}^{-1}}\right)^{-1}(1-\epsilon),
\end{equation} 
{where $T_{\rm diss}$ is the lifetime of the cluster, $N$ is the
  number of cluster stars, $\rg$ the galactocentric distance, $\vg$
  the circular velocity of the galactic model, and $\epsilon$ the
  eccentricity of the orbit. Then we use the parameterization for the
  evolution of the MF slope as a function of the time elapsed until
  cluster dissolution}

\begin{equation}
\label{slope2}
\alpha = \alpha_0 - 1.51\left(\frac{T}{T_{\rm diss}}\right)^2 + 1.69\left(\frac{T}{T_{\rm diss}}\right)^3 - 1.5\left(\frac{T}{T_{\rm diss}}\right)^4.
\end{equation}

Under the assumption that the cluster has moved in circular orbit, at
the solar circle $\rg = 8\,$kpc, for 8$\,$Gyr, equation
(\ref{eq:lifetime}) gives a dissolution time of $T_{\rm diss} \approx
12\,$Gyr; taking this value as input for equation (\ref{slope2}), the
predicted slope of the present-day MF at the low-mass end is $\alpha
\approx 0.83$, which corresponds to a MF significantly steeper than
the observed one ($\alpha \approx 0.2$).

{In this way the result from the parameterization of the MF slope
  \citep{2003MNRAS.340..227B} applied to the observational data of
  \citet{2005AJ....130..626K} is a validation of the initial mass we
  infer for NGC~6791. Also, showing that the observed MF is flatter
  than what is expected for a cluster that has always been at 8$\,$kpc
  supports our hypothesis that NGC~6791 experienced stronger tides in
  the past.}

\subsection{Velocity dispersion}

Regarding the velocity dispersion, \citet{2006ApJ...643.1151C} obtain
radial velocities for a subsample of 15 giant stars in NGC~6791 and
report a radial velocity dispersion of $\sigma_r = 2.2 \pm 0.4\kms$. More
recently, \citet{2011ApJ...733L...1P} note that if a King model is
assumed to describe the inner mass profile of NGC~6791, then the
projected velocity dispersion averaged within $\rh$ should be
$\sigma_r \simeq 0.75 \kms$, a value significantly lower than the one
reported by \citet{2006ApJ...643.1151C}.

\begin{figure}
\begin{center}
\includegraphics[width=9cm]{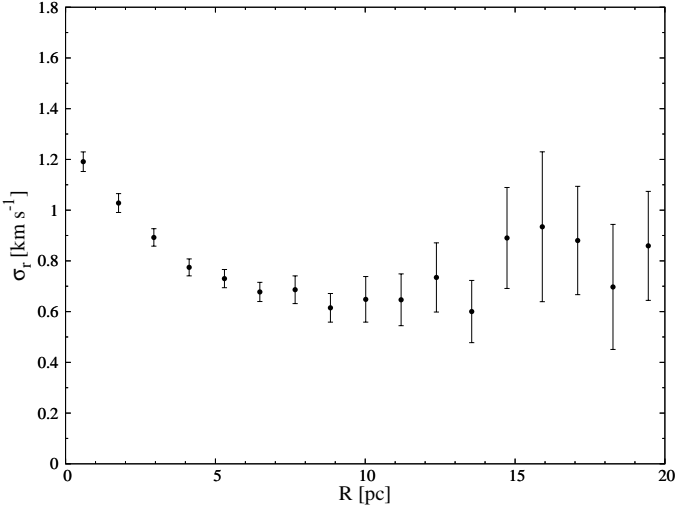}
\end{center}
\caption{Projected velocity dispersion as a function of projected
  radius within the cluster.}
\label{f10}
\end{figure}

In Figure \ref{f10} we compute the projected velocity dispersion as a
function of radius within the cluster, for the model at
$7.6\,$Gyr. Notice that the dispersion in projected velocities is
small near the cluster centre, and increases at the boundaries, with a
peak value near the tidal radius. Hence, in our models we found that
$\sigma_r$ inside the half mass radius is small, similar to the
expected value from \citet{2011ApJ...733L...1P}, and much lower than
the one reported by \citet{2006ApJ...643.1151C}. The high dispersion
measured by Carraro et al. may be due to orbital motions of binary
stars that increase the velocity dispersion, and which are not
included in our $N$-body model, nor in the (virial) estimate of
Platais et al.

From Figure \ref{f10} we can see two different behaviours; in the
central parts of the cluster $\sigma_r$ systematically decreases with
radius, however, for larger radii, it remains constant or even
increases. This increase is most likely due to the aforementioned
potential escapers \citep{2000MNRAS.318..753F, 2017MNRAS.466.3937C}.

\section{The disruptive effect of interactions with giant molecular clouds }
\label{sec:gmcs}
In our $N$-body model we did not include molecular gas, and it is
well-known that interactions with giant molecular clouds (GMCs) are an
efficient cluster disruption mechanism
\citep{1958ApJ...127...17S,2006MNRAS.371..793G}. Here we quantify the
disruptive effect of the Milky Way disc by using a simple model for
the gas distribution in the Milky Way and by following one of our
cluster orbits through the gas.

\citet{1958ApJ...127...17S} showed that the disruption timescale as
the result of impulsive tidal shocks from passing GMCs can be
expressed as

\begin{equation}
\taugmc = \ggmc \rhoh,
\label{eq:taugmc}
\end{equation}
where $\rhoh$ is the average density within the half-mass radius of
the cluster, and $\ggmc$ depends on the properties of the molecular
gas as

\begin{equation}
\ggmc \propto \frac{\sigmagmc}{\Sigmagmc\rhoism},
\end{equation}
with $\sigmagmc$ is the dispersion of the relative velocities between
the cluster and the GMCs, $\Sigmagmc$ is the surface density of
individual GMCs and $\rhoism$ is the volume density of cold molecular
gas.

We consider a density profile for the cold molecular gas of the form

\begin{equation}
\rhoism(R_{\rm GC},z) = \rho_0\exp\left(-\frac{R_{\rm GC}}{L}\right)\sech^2\left(-\frac{Z}{2h}\right).
\end{equation}
Here $R_{\rm GC} = \sqrt{X^2 + Y^2}$ is the distance to the Galactic
centre when in the midplane and $Z$ is the height above the disc. The
(vertical) velocity dispersion is $\sigmaz(R_{\rm GC}) = h\sqrt{8\pi
  G\rho_0\exp(-R_{\rm GC}/L)}$ and is independent of $Z$. We use a
scale length of $L = 2.5\,$kpc and a scale height of $h=60\,$pc
\citep{2008gady.book.....B} and we chose $\rho_0=0.74\,\msun/\pc^3$,
such that the mid-plane density at the solar radius ($R=8\,$kpc)
equals $0.03\,\msun/\pc^3$ \citep{1987ApJ...319..730S}.

We follow a particle orbit through this disc and determine the quantity 
\begin{equation}
f = \left(\frac{\int_0^{\rm 8\,\gyr}\ggmc^\odot/\ggmc(t^\prime)\dr t^\prime}{8\,\gyr} \right)^{-1}
\end{equation}
where $\ggmc^\odot$ is the quantity in equation~(\ref{eq:taugmc}) for
the disruption in the mid-plane in the solar neighbourhood. For the
orbit described in section \ref{sec:mass} we find $f \simeq
2.3$. \citet{2016MNRAS.463L.103G} showed that because of the
self-limiting nature of tidal shocks, the timescale for dissolution
scales as $\ggmc^{1/3}$. Taking the harmonic time average of this
quantity, we find $f\simeq1.9$.
 
From this exercise we find that, despite the fact that NGC~6791 formed
near the Galactic centre, the average effect of disruption by GMCs
during its life is less important than in the mid-plane at the solar
radius. This can be understood by the large vertical excursions form
the disc, where the GMC density is low.

Using the results of \citet{2016MNRAS.463L.103G}, we estimate that the
(average) disruption time-scale by GMCs is $\sim6\,\gyr$ such that the
initial mass of NGC\,6791 could have been a factor of $\sim2$ higher
than what we derived in Section~\ref{sec:Nbody} if we had included
interactions with the molecular gas. The initial mass could not have
been much higher, otherwise the MF would have been even flatter.

\section{Discussion}
\label{discussion}

\subsection{Plausibility of radial migration}
Several papers have studied the orbit of NGC~6791, trying to test the
plausibility of a radial migration scenario that displaced the orbit
of the cluster from the inner Galaxy to its present-day location
\citep{2006ApJ...643.1151C,2006A&A...460L..27B,2009MNRAS.399.2146W,2012A&A...541A..64J}. Most
of these works integrate orbits within galactic potentials that are
either axisymmetric or incorporate bar and spiral arms in a rather
simplified way (a potential that does not try to reproduce an existing
mass distribution). It is known that in order for radial migration to
happen, the orbit should modify its angular momentum, a process not
possible within an axisymmetric potential. In this respect, by
including a bar and spiral arms, \citet{2012A&A...541A..64J} present
an improvement over previous works, however they still treat the
spiral arms as a perturbation not based in physical overdensities,
which underestimates their dynamical effect \citep{PMME03}; {in
  addition to this; the authors are missing an extra velocity
  transformation that is key for their study, so we should caution
  making conclusions about the orbit of NGC~6791 based on their
  analyses.

Although some authors argue that the radial migration process is not
efficient enough to move the cluster more than a couple of kpc, nor
able to lift it to its present position \citep{2012A&A...541A..64J,
  2017ApJ...842...49L}, in this work, employing a Galactic model based
on physical mass distributions of a bar and detailed spiral arms, we
find that a fraction of the newly formed clusters will eventually (8
Gyr) migrate more than 4 kpc and be lifted more than 800 pc away from
the disc mid-plane; specifically, we found abundant examples of orbits
that match the current PMs and high altitudes of NGC~6791 that started
their evolution in the inner thin disc.

Due to the dynamical chaotic nature of the system (produced by the
interactions with the bar and the spiral arms) we find hundreds of
distinct orbits that fulfill our selection criteria (i.e. orbits whose
present day orbital parameters correspond to those observed in NGC
6791); such orbits have very different initial conditions and very
different overall trajectories, and it is therefore not possible to
define an exact initial position, nor the amount of encounters
experienced during the orbit.

\subsection{Evidence supporting the formation in the inner Galaxy}

The evidence that links the formation place of NGC 6791 to the inner
Galaxy comes from different observables: It is one of the most
metal-rich open clusters in the Galaxy, $[{\rm Fe/H}]\sim+0.40$. Stars
with the age of NGC~6791 and with similar $\feh$ are not present at
the galactocentric distance of the cluster, but are found in the inner
thin disc and in the bulge, $3\,$kpc $< R_0/\kpc < 5$. More recently,
\citet{2017ApJ...842...49L}, using the APOGEE DR13 dataset, placed
NGC~6791 in the $[\alpha/{\rm Fe}] \,{\it vs.}\, [{\rm Fe/H}]$ plane;
the cluster's position on this plane strongly suggests a connection
with the MW bulge as well as with the inner thin disc. However, Linden
et al. dismiss an origin in the thin disc arguing that it is unlikely
for a radial migration mechanism to operate by several kpc, and
especially to account for the current cluster's high altitude above
the plane. Although such arguments could prove that it is unlikely
that any single cluster would migrate and raise so much, when there
are thousands of clusters that are born, it becomes a fact that a few
of them will migrate and raise just as much, here we found that the
mechanism exists and can reproduce the current orbit of NGC~6791.

On the other hand, HST observations reveal a rather flat stellar mass
function in the central regions of NGC 6791
\citep{2005AJ....130..626K}. It is expected that the MF in a stellar
cluster flattens with time due to the steady tidal stripping of mainly
the low mass stars. However, the mass loss of the cluster, by
removing these stars, must be very efficient in order to achieve a
totally flat MF.

Regarding the mass loss, \citet{2015MNRAS.449.1811D} find signatures
of tidal distortions in the density distribution of NGC 6791,
suggesting that the cluster is still experiencing mass loss. Based on
this evidence, the authors argue that, at some point during its
evolution, NGC 6791 may have lost an important fraction of its
original mass.

Notice that the observed tidal distortions and flat MF require a
significant mass loss to be explained, further supporting the scenario
in which NGC 6791 was formed in the inner thin disc or in the bulge
(where the tidal field is strong), and undermining the suggestion that
it was formed in the nearby thick disc (where the tidal field is
much weaker). However, these two observed properties are not enough to
restrict the formation place of NGC 6791 because a significant mass
loss can also be achieved during violent encounters of the cluster
with different Galactic structures at any point along its orbit.

Nonetheless, the high metallicity of NGC~6791 is a key property and
the clearest clue about its origin. Its value, along with its age,
places the cluster's birth at the inner thin disc or in the bulge, $3
< R_0/\kpc < 5$. In this work we show that it is possible that the
cluster migrated from there to its current orbit, and survived.

\section{Conclusions}
\label{conclusions}
With the use of an observationally motivated, three dimensional
Galactic mass model, we perform a comprehensive orbital study of the
enigmatic open cluster NGC~6791. We integrate half million orbits,
representing the MW stellar disc, for 8 Gyr to find those with similar
position, proper motion, and radial velocity to the current values of
NGC~6791.

Contrary to previous results that claim that radial migration is
inefficient in moving the cluster to its present position
\citep[e.g.][]{2012A&A...541A..64J, 2017ApJ...842...49L}, we find
  240 examples of orbits that match the current position and velocity
  of NGC~6791. We expect that out of every 420 massive clusters that
  were born between 7.5 and 8.5 Gyr ago, with a galactocentric radius
  between 3 and 5 kpc, one would survive to this day, be far away from
  the plane, be anomalously metal rich, and be observable from
  Earth. I.e. there is a probability that an orbit could have
suffered an outward radial migration that brought the cluster to its
current galactocentric position and its current distance from the
plane. This scenario would explain its high metallicity, in spite of
its old age.

As a consequence of the previous point, to speak of an exact initial
position, or of specific encounters suffered by the orbit of NGC~6791,
becomes meaningless in view of its highly chaotic nature produced by
the interactions with the bar and the spiral arms.

In order to test the survival of a cluster within the found family of
orbits that matches NGC~6791 orbital parameters, we perform two
studies, one based on a self-consistent field technique and another
based on direct $N$-body models including the time-dependent Galactic
tides and the effects of stellar evolution.

The Galactic environments the cluster went through, play a key role in
its evolution, it determines the possible initial mass of
NGC~6791. This implies that, unless we know the exact orbit, the mass
cannot be uniquely determined. Nevertheless, in order to endure
violent Galactic environments, the cluster should have been born
significantly more massive ($\sim5\times10^4\,\msun$) than it is today
($\sim5\times10^3\,\msun$).

As a byproduct of an efficient early tidal stripping, achievable in
the strong tidal field of the inner Galaxy, our simulations reproduce
a flat stellar mass function, as observed for NGC~6791. Therefore, the
birth place and journeys of NGC~6791 are imprinted in its chemical
composition, in its mass loss, and in its flat stellar mass function,
further confirming its origin in the inner thin disc or in the bulge.

\section*{Acknowledgements}
We thank Lucie J{\'{\i}}lkov{\'a} and Giovanni Carraro for discussions on their model and in particular for redoing their orbit calculations with the updated velocities which brought their results in better agreement with our findings (see Section~\ref{analysis}).
We would like to thank Florent Renaud and Emanuele Dalessandro for
fruitful discussions, and an anonymous referee for an insightful
review of this work. We are also grateful to Sverre Aarseth and Keigo
Nitadori for making {\sc nbody6} publicly available. We acknowledge
DGTIC-UNAM for providing HPC resources on the Cluster Supercomputer
Miztli. L.A.M.M and B.P. acknowledge CONACYT Ciencia B\'asica grant
255167 and DGAPA PAPIIT IG 100115. A.P. acknowledges DGAPA-PAPIIT
through grant IN-109716. L.A.M.M. acknowledges support from DGAPA-UNAM
postdoctoral fellowship. M.G. acknowledges financial support from the
Royal Society (University Research Fellowship) and the European
Research Council (ERC StG-335936, CLUSTERS). L.A.M.M. thanks the ERC
for partially funding a visit to Surrey where most of this work was
done.


\end{document}